\documentclass[preprint, showpacs,amsmath,amssymb]{revtex4}
\usepackage{graphicx}

\begin{document}

\title{Quantitative broadband chemical sensing in air-suspended solid-core fibers}

\author{T. G. Euser}
\affiliation{Max-Planck Research Group (IOIP), University of Erlangen-Nuremberg, G\"unther-Scharowsky-Str. 1/Bau 24, 91058 Erlangen, Germany.}
\email{teuser@optik.uni-erlangen.de}

\author{J. S. Y. Chen}
\affiliation{Max-Planck Research Group (IOIP), University of Erlangen-Nuremberg, G\"unther-Scharowsky-Str. 1/Bau 24, 91058 Erlangen, Germany.}

\author{N. J. Farrer}
\affiliation{Department of Chemistry, University of Warwick, CV4 7AL, Coventry, United Kingdom.}

\author{M. Scharrer}
\affiliation{Max-Planck Research Group (IOIP), University of Erlangen-Nuremberg, G\"unther-Scharowsky-Str. 1/Bau 24, 91058 Erlangen, Germany.}

\author{P. J. Sadler}
\affiliation{Department of Chemistry, University of Warwick, CV4 7AL, Coventry, United Kingdom.}

\author{P. St.J. Russell}
\affiliation{Max-Planck Research Group (IOIP), University of Erlangen-Nuremberg, G\"unther-Scharowsky-Str. 1/Bau 24, 91058 Erlangen, Germany.}

 \homepage{http://www.pcfibre.com}

\date{Manuscript prepared on December 14th 2007.}

\begin{abstract}
We demonstrate a quantitative broadband fiber sensor, based on evanescent field sensing in the cladding holes of an air-suspended solid-core photonic crystal fiber. We discuss the fabrication process, together with the structural- and optical characterization of a range of different fibers. Measured mode profiles are in good agreement with finite element method calculations made without free parameters. The fraction of the light in the hollow cladding can be tuned via the core diameter of the fiber. Dispersion measurements are in excellent agreement with theory and demonstrate tuning of the zero dispersion wavelength via the core diameter. Optimum design parameters for absorption sensors are discussed using a general parameter diagram.  From our analysis, we estimate that a sensitivity increase of three orders of magnitude is feasible compared to standard cuvette measurements. Our study applies to both liquid and gas fiber sensors. We demonstrate the applicability of our results to liquid chemical sensing by measuring the broad absorption peak of an aqueous NiCl$_2$ solution. We find striking agreement with the reference spectrum measured in a standard cuvette, even though the sample volume has decreased by three orders of magnitude. Our results demonstrate that air-suspended solid-core PCFs can be used in quantitative broadband chemical sensing measurements.
\end{abstract}

\maketitle

\section{INTRODUCTION}
\label{introduction}

Recent advances in photonic crystal fiber (PCF) \cite{russell2003} design have generated much interest in exploring the use of PCFs as vehicles for new kinds of optical sensors and devices. Fibers in which a large fraction of the guided mode propagates in the hollow region of the microstructure, such as hollow-core PCF (HC-PCF),\cite{cregan1999} are exploited for their ability to maximize the interaction of light and sample at path lengths that are much longer than achievable in conventional sample cells. At the same time, PCF sensors can strongly reduce the sample volume required for measurements and provide the robustness and flexibility needed for fiber sensors.

In HC-PCF, the guiding mechanism is based on the photonic bandgap in the cladding structure, which keeps the propagating light trapped and guided in the hollow core of the fiber, thus offering an ideal environment for optical spectroscopy. While HC-PCF sensors are highly suitable for narrow-band spectroscopic gas sensing measurements,\cite{benabid2005,kornaszewski2007} HC-PCF sensors are unable to resolve broad spectral features, as their transmission bandwidth is typically narrower than 100~nm. In this respect, HC-PCF sensors are not competitive with the current commercial table-top UV-VIS spectrometers.

To overcome the limited bandwidth of HC-PCFs, a different class of PCFs, known as the solid-core index-guiding PCFs, has been proposed and demonstrated in the literature.\cite{jensen2004,cordeiro2006} Here, the guiding mechanism is based on the index contrast between the solid core and the microstructured cladding of the fiber, analogous to that of the conventional optical fiber. The most common design of an index-guiding PCF consists of a solid silica core surrounded by a periodic array of silica webs and air holes that make up the cladding. Unlike the HC-PCF, which offers direct light-sample interactions, the propagating light in the solid core of the index-guiding PCF probes the sample via an evanescent field. By manipulating the core size and the pitch of the cladding air holes the amount of evanescent field available for sensing the sample present in the cladding holes can be varied. However, the maximum achievable fraction of power overlap $\Phi$ in the cladding holes in these fibers is usually too low (less than 10\%) to allow ultra high sensitivity sensing.

An index-guiding PCF design that strongly enhances the power overlap in the cladding holes consists of a solid silica rod held in air by three silica nanowebs. This allows direct access to the fiber core for sensing applications.\cite{monro1997} By varying a single structural parameter the fraction of evanescent field available for sensing can be controlled while maintaining the broad transmission window of silica. Recently, narrow spectral lines of acetylene were resolved using the evanescent field of light propagating in such fibers.\cite{webb2007,cordeiro2007} So far, sensing experiments in these PCFs to date have been limited to a narrow frequency range, but this fails to exploit the full potential of these fibers.

In this article we report a quantitative broadband in-fiber sensor, that uses evanescent field-sensing in air-suspended solid-core PCFs. We discuss the fabrication process and the full characterization (both theory and experiment) of a range of different fibers. We discuss general design parameters for evanescent field sensors and demonstrate the potential of our fiber sensor by a quantitative measurement of the broad absorption peak for an aqueous NiCl$_2$ solution. We find striking agreement with a reference spectrum measured in a standard cuvette, demonstrating that this fiber can be used in quantitative broadband chemical-sensing measurements.

\section{FABRICATION}
\label{fabrication}

The air-suspended solid-core PCFs were fabricated using the conventional stack-and-draw process. Silica tubes were first drawn to thin-walled capillaries of the desired diameter. Three of these capillaries were stacked and inserted into a jacket tube to form the preform of the fiber. This simple preform illustrates another advantage of this fiber over HC-PCF, the preform of which typically contains over 300 capillaries and rods. The preform was subsequently drawn directly into fiber. During drawing, the core size of the fiber, which determines the amount of evanescent field available for sensing, can be controlled via the scale-down ratio of the perform. Using this technique we have fabricated kilometers of fibers with 10 different core diameters in the range of 0.82 to 3.0~$\mu$m.

\begin{figure}[!ht]
\begin{center}
\includegraphics[width=\linewidth]{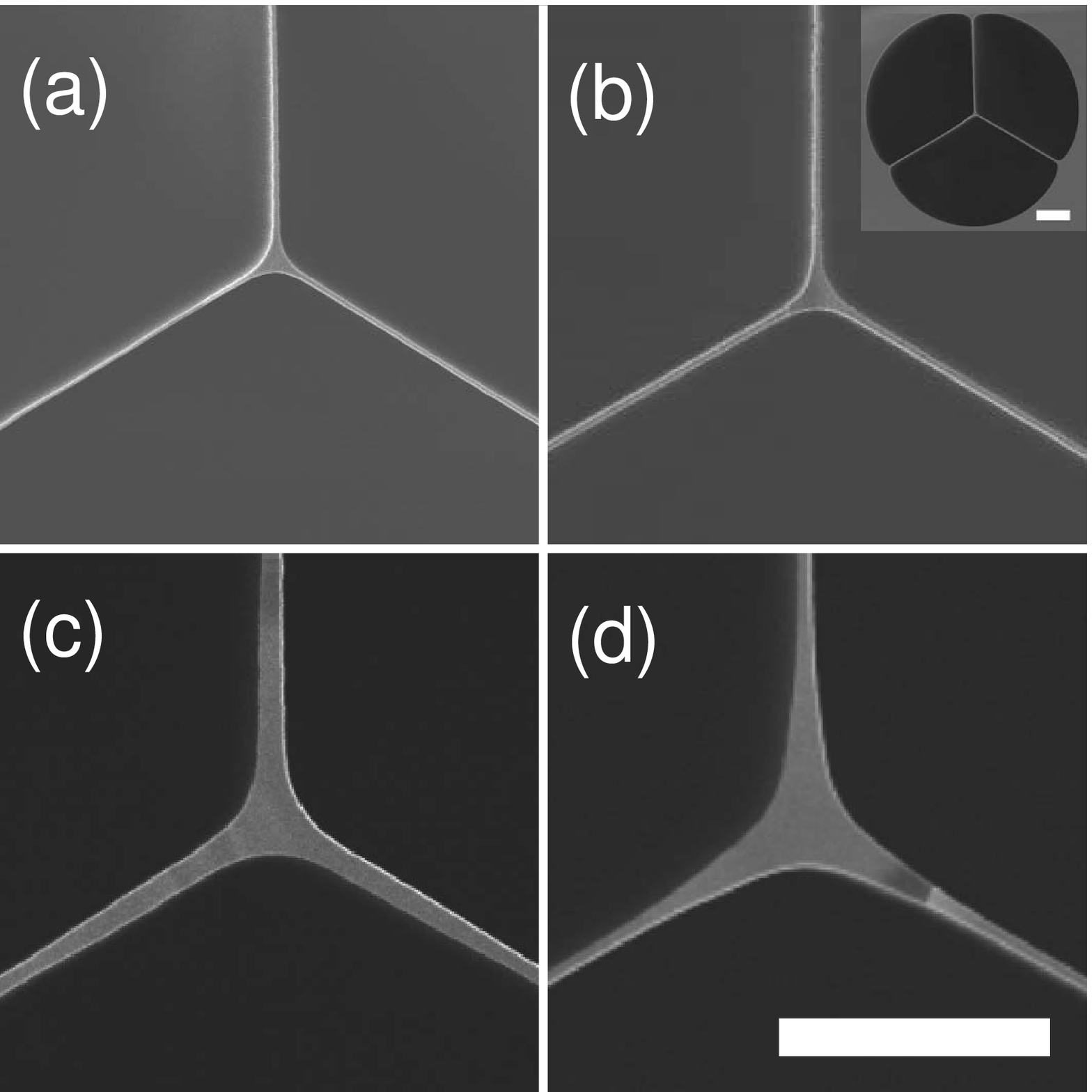}\\
\caption{\label{fig-SEM}High resolution scanning electron microscope images of the core region of four different air-suspended solid-core fibers. The core diameter $d_{\mathrm{eff}}$ of the fibers is defined as the diameter of the largest circle that can be drawn in the core region. We deduce $d_{\mathrm{eff}}$=0.87~$\mu$m (a), 1.03~$\mu$m (b), 2.32~$\mu$m (c) and 2.98~$\mu$m (d), for Fibers 1 to 4, respectively. The inset in (b) shows the hollow cladding region of Fiber 2 with a diameter of 64~$\mu$m. The thicknesses of the nanowebs that hold the core in place vary between 160 to 500~nm. Structural variations along the fibers used in our experiments were measured to be less than 2\% over tens of meters. Both scalebars represent 10~$\mu$m.}
\end{center}
\end{figure}

Figure~\ref{fig-SEM} shows typical scanning electron microscope (SEM) images of the fibers drawn. Three nanowebs with thicknesses between 160~nm and 550~nm hold the central silica core in place. We define the effective core diameter $d_{\mathrm{eff}}$ of the air-suspended solid-core PCF as the diameter of the largest circle which can be inscribed in the core region. For the fibers shown in Fig.~\ref{fig-SEM}, the core diameters are 0.873, 1.03, 2.32 and 2.98~$\mu$m. The hollow cladding region (see the inset in Fig.~\ref{fig-SEM}(b)) acts as an easily accessible sample chamber and has a typical diameter of 30-70~$\mu$m. The consistency of the structural parameters along the fiber was verified from high resolution SEM images. For most samples variations of less than 3.5\% over tens to one hundred meters were detected. The largest variation in $d_{\mathrm{eff}}$ observed was less than 7\%. For the fibers used in our experiments the observed variations were below 2\%. Our analysis demonstrates that the stack-and-draw process allows highly reproducible and flexible fabrication of air-suspended solid-core PCFs.

\section{TRANSMISSION AND LOSSES}
\label{tx-loss}

\begin{figure}[!ht]
\begin{center}
\includegraphics[width=\linewidth]{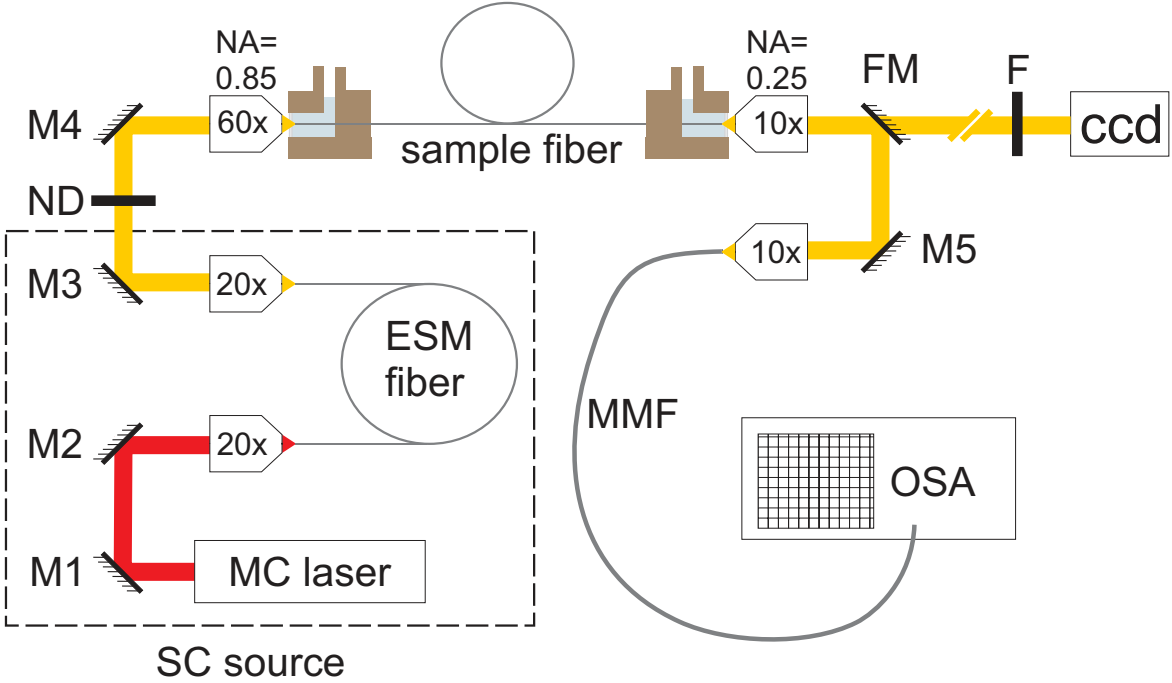}\\
\caption{\label{fig-setup}
(Color online) Schematic of the setup. The dashed rectangle indicates the supercontinuum (SC) source (see Ref.~\onlinecite{birks1999}). Light from a 1064~nm microchip (MC) laser which produces 4~$\mu$J, 1-2~ns pulses at a repetition rate of 40~kHz is launched into the core of a 20~m long endlessly single mode (ESM) fiber to generate a broadband SC. The SC light is collimated by a 20$\times$ objective, attenuated by a neutral density filter (ND), and launched into the core of our sample fiber by a 60$\times$0.85NA objective. Both fiber ends are mounted into water cells, facilitating the flow of sample liquid through the cladding holes. The transmitted light is picked up by a 10$\times$0.25NA objective. The light from the core mode can be projected onto a CCD beam profile camera (CCD), positioned 2~m away from the fiber. Interference filters (F) are used to select the desired wavelength range. A flip mirror (FM) is used to couple the light into a multimode fiber (MMF) which is connected to a Yokogawa AQ-6315A optical spectrum analyzer (OSA). M1-M5 are mirrors.}
\end{center}
\end{figure}

The transmission and loss spectra are important because they provide guidelines to the wavelength range and maximum fiber length that can be used in sensing experiments. They were determined using a broadband supercontinuum (SC) source with emission in the wavelength range from  $\lambda$=480~nm to beyond $\lambda$=1750~nm. The SC was generated from an endlessly single mode PCF (ESM-PCF) pumped by a Q-switched Nd:YAG microchip laser at $\lambda$=1064~nm.\cite{wadsworth2004} The generated SC is coupled into the sample fiber via an objective, allowing more freedom over coupling parameters than fiber butt-coupling. A schematic of the setup is shown in Fig.~\ref{fig-setup}.

For the experiments in liquid-filled fibers, both fiber ends were connected to custom-made water cells to allow complete infiltration of liquid solution into the cladding holes of the fiber. The sample cell design also allows for continuous flow measurements. The water cells have a thin glass window which allows the use of high NA objectives for coupling of light in and out of the fiber. The dead volume in the water cells is limited to only 50~$\mu$L, they can resist pressures of up to 10 bar. The cladding holes have a diameter of typically 30~$\mu$m, allowing rapid filling of the sample using pressures of the order of 1 bar (typically within 1-2 seconds for a 1 m long fiber). In our experiments we have used a standard syringe to fill the cladding holes of the fibers.

The coupling of light into the fibers is optimized by matching the numerical aperture (NA) of the coupling objective to that of the fiber. The NA of the fiber can be approximated by NA=sin($\sqrt{n_1^2-n_2^2}$), where $n_1$ and $n_2$ are the refractive indices of the core and cladding material respectively. For a water-filled fiber at $\lambda$=700~nm, we find  NA=0.58. After correcting for the change from objective (air) to fiber (water) we obtain the effective NA of the fiber/water-cell combination to be NA$_{\mathrm{eff}}$=0.58*(1.33/1)=0.77. To optimize the amount of light that is coupled into the fiber, we used a 60$\times$0.85NA objective whose NA accords with NA$_{\mathrm{eff}}$. A CCD beam profiler imaging the output-facet of the fiber was used to verify that light was coupled only into the fundamental core mode.

\begin{figure}[!ht]
\begin{center}
\includegraphics[width=\linewidth]{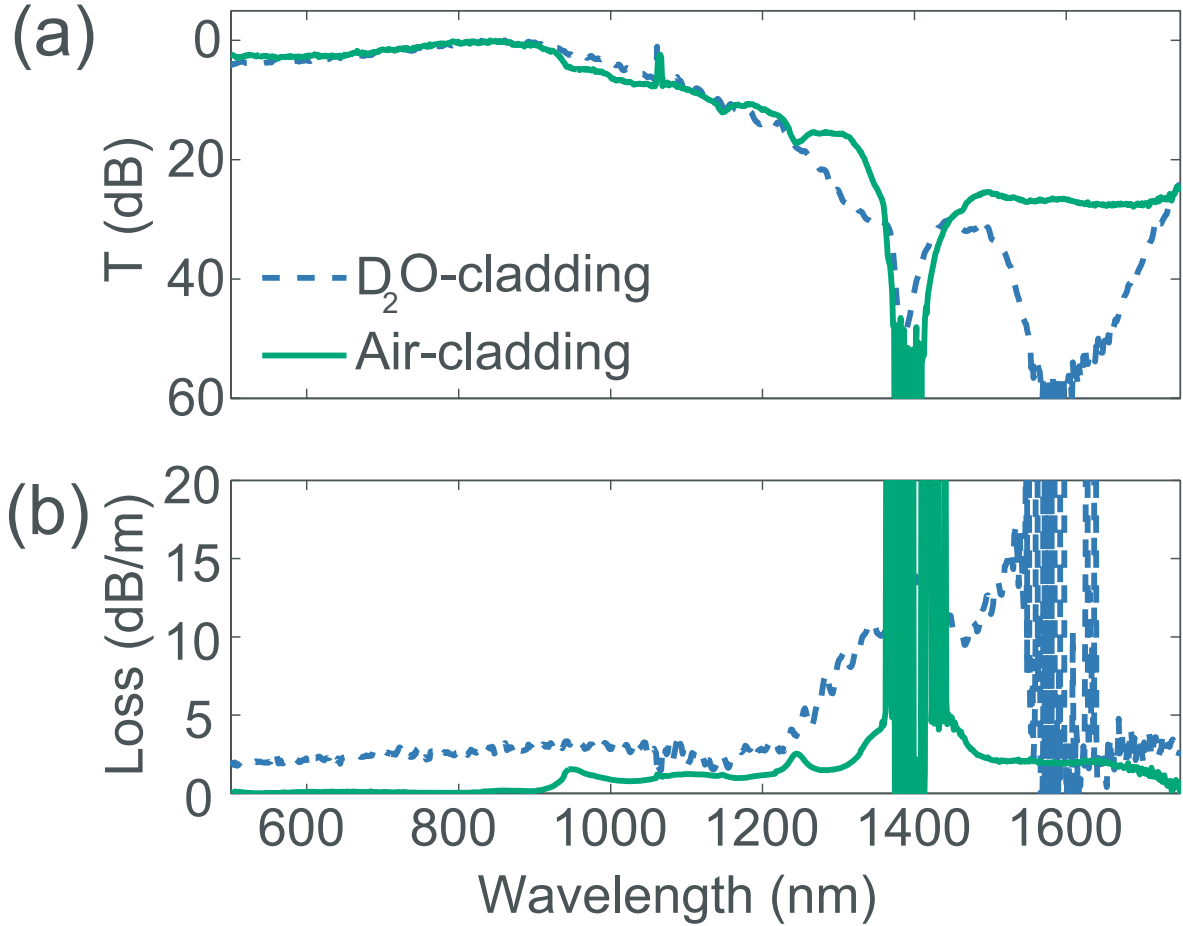}\\
\caption{\label{fig-tx-loss}(Color online) Transmission normalized to SC source T (a) and fiber loss spectra (b) for 2.9~m of Fiber 2 with air cladding (solid green curves) and 1.0~m of Fiber 3 infiltrated with D$_2$O (dashed blue curves). Both fibers display transmission between $\lambda$=500 and 1750~nm, apart from a narrow absorption band around $\lambda$=1400~nm, which is due to a residual H$_2$O contamination of the silica. Transmission of the D$_2$O-filled fiber displays an additional absorption band around $\lambda$=1600~nm, which agrees with the expected shift of the absorption band of D$_2$O for this region. The loss spectra (b) show low losses $<$0.2~dB/m for Fiber 2 with air cladding between $\lambda$=480 and 900~nm. The maximum loss within the transmission windows is 4~dB/m. Losses of the D$_2$O-filled Fiber 3 remain below 3~dB/m between $\lambda$=480 and 1220~nm.}
\end{center}
\end{figure}

At the output-facet of the fiber a 10$\times$0.25NA objective was used to collimate the transmitted light. A flip mirror allowed us to send the light either to a CCD beam profiler (see Section~\ref{fig-mode-profile}) or to a multimode fiber that is connected to an optical spectrum analyzer (OSA). The transmission spectra were normalized to the SC reference spectrum. Figure~\ref{fig-tx-loss}(a) shows the broadband transmission windows for 2.9 m of Fiber 2 with air cladding from $\lambda$=480 to 1350~nm and from $\lambda$=1450 to beyond 1750~nm. The absorption line near $\lambda$=1400~nm is attributed to H$_2$O contamination during the fiber drawing process. This H$_2$O absorption can be reduced drying the silica perform or by treating it with chlorine gas. The dashed curve shows the transmission window for 1 m of Fiber 3 infiltrated with heavy water (D$_2$O). The refractive index of D$_2$O is similar to that of H$_2$O. However,
all absorption bands are shifter to longer wavelengths due to the almost doubled moment of inertia of D$_2$O compared to H$_2$O, which reduces the vibrational frequencies by a factor of about $\sqrt{2}$.  The measured transmission of the D$_2$O-filled fiber indeed displays an absorption band at $\lambda$=1600~nm, which accords with the H$_2$O-absorption band at 1190 nm, shifted by $\sim\sqrt{2}$.\cite{hale73} The transmission spectra show that the fibers guide light over a broad wavelength range, allowing sensing measurements between $\lambda$=480 and 1750~nm.

The losses of all fibers were determined via the conventional cut-back technique,\cite{marcuse1981} in which the power transmitted through a long length of fiber is measured and normalized to the power transmitted through a shorter length of the same fiber without changing the in-coupling condition. A typical resulting loss spectrum for Fiber 2 with air cladding is shown in Fig.~\ref{fig-tx-loss}(b). The spectrum reveals a low-loss region with losses below 0.2~dB/m between $\lambda$=480 and 900~nm. The maximum observed loss within the transmission windows of this fiber is 4~dB/m. The loss spectrum was also measured in a D$_2$O-filled Fiber 3. The losses are slightly higher than in the air-filled Fiber 2 but remain below 3~dB/m over the range between $\lambda$=480 and 1220~nm. For typical sensing measurements, fiber lengths of less than 20~m suffice. We therefore conclude that losses do not limit the performance of these fiber sensors. The loss spectra for both the air-filled and H$_2$O-filled fibers are flat over a broad wavelength range, which leads to the conclusion that the length of a fiber sensor can be adjusted without changing the shape of the transmission spectra.

\section{MODE FIELD DISTRIBUTION}
\label{mode-field}

\begin{figure}[!ht]
\begin{center}
\includegraphics[width=\linewidth]{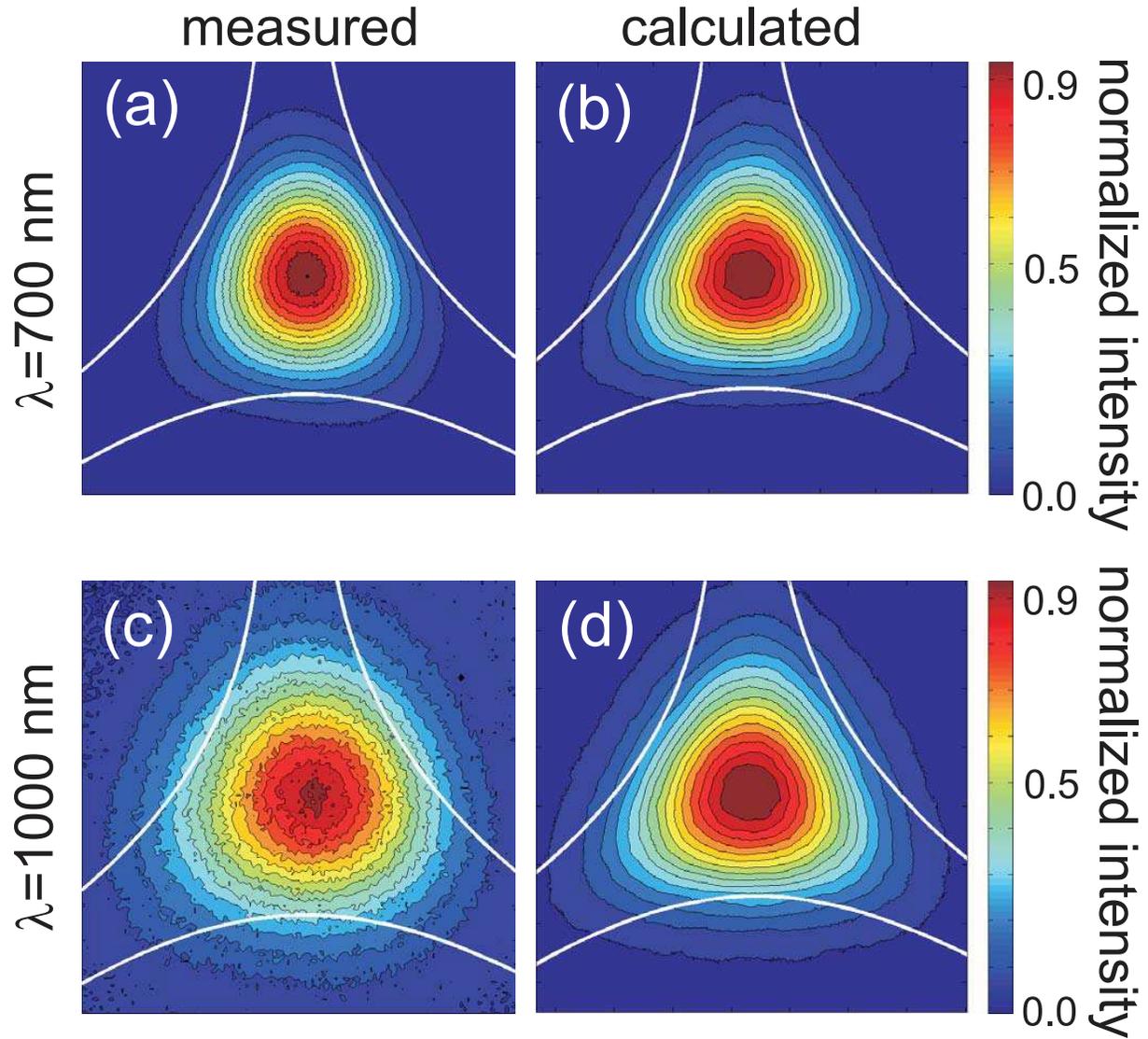}
\caption{\label{fig-mode-profile}(Color online) Normalized mode profiles of the Fiber 2 with  H$_2$O-filled cladding. All images are 2$\mu$m$\times$2~$\mu$m; the white curves indicate the contours of the fiber structure. (a) and (b) show measured and calculated mode profiles respectively, at $\lambda$=700~nm. The fraction of the light that extends into the cladding holes is $\Phi\sim$10\%. (c) and (d) show measured and calculated mode profiles respectively, at $\lambda$=1000~nm. The fraction of the light that extends into the cladding holes increases to $\Phi$=20\%}
\end{center}
\end{figure}

The sensing mechanism in air-suspended solid-core PCFs is based on the overlap between the evanescent field of the guided mode and the sample. In quantitative sensing experiments it is essential to know the fraction of power $\Phi$ in the cladding holes that is available for interaction with the sample. We have therefore measured the mode profiles of our fibers at various wavelengths. Measurements of the near-field images of the guided mode were taken with a CCD beam profiler (WinCamD-UHR-1310) by imaging the output facet of the fiber onto the CCD with a 60$\times$0.85NA microscope objective. By placing the CCD camera at a distance of about 2 m from the imaging objective, we ensured that only the core region was imaged by the beam camera. We calibrated the scale of the beam camera by translating the fiber stage over known distances. The resulting beam profiles for H$_2$O-filled Fiber 2 at two different wavelengths are shown in Figs.~\ref{fig-mode-profile}(a) and (c). The measured profiles show that at shorter wavelengths $\lambda$=700~nm (Fig.~\ref{fig-mode-profile}(a)), the mode is confined to the core. As $\lambda$ increases to 1000~nm (Fig.~\ref{fig-mode-profile}(a)) the field extends further into the cladding holes, implying that a higher power fraction in the cladding holes is available for sensing.

The mode profiles were also calculated using the finite element method (FEM). The calculations were based on contours extracted from SEM images of the measured fibers and consequently do not contain any freely adjustable parameters. The fiber structure is discretized using triangular elements of 0.05~$\mu$m in the core region and larger elements in the cladding region to achieve a realistic discretization of the fiber structure. Figures~\ref{fig-mode-profile}(b) and (d) show the calculated mode profiles of the fibers for various core sizes and cladding media at different wavelengths. The overall shapes of the calculated profiles agree well with the measured profiles.

Some profiles calculated using the FEM revealed discontinuities on a sub-100~nm scale across the glass-air (core-cladding) boundary. These discontinuities are attributed to field enhancement, caused by the discontinuity in the normal electric field given by the ratio between the dielectric constants of the two media.\cite{wiederhecker2007} These near-field features do not appear in our measured beam profiles since they have dimensions that are well below the free space diffraction limit of the light. Scanning near-field optical microscopy can be used to resolve such  features.\cite{wiederhecker2007}

\begin{figure}[!ht]
\begin{center}
\includegraphics[width=0.8\linewidth]{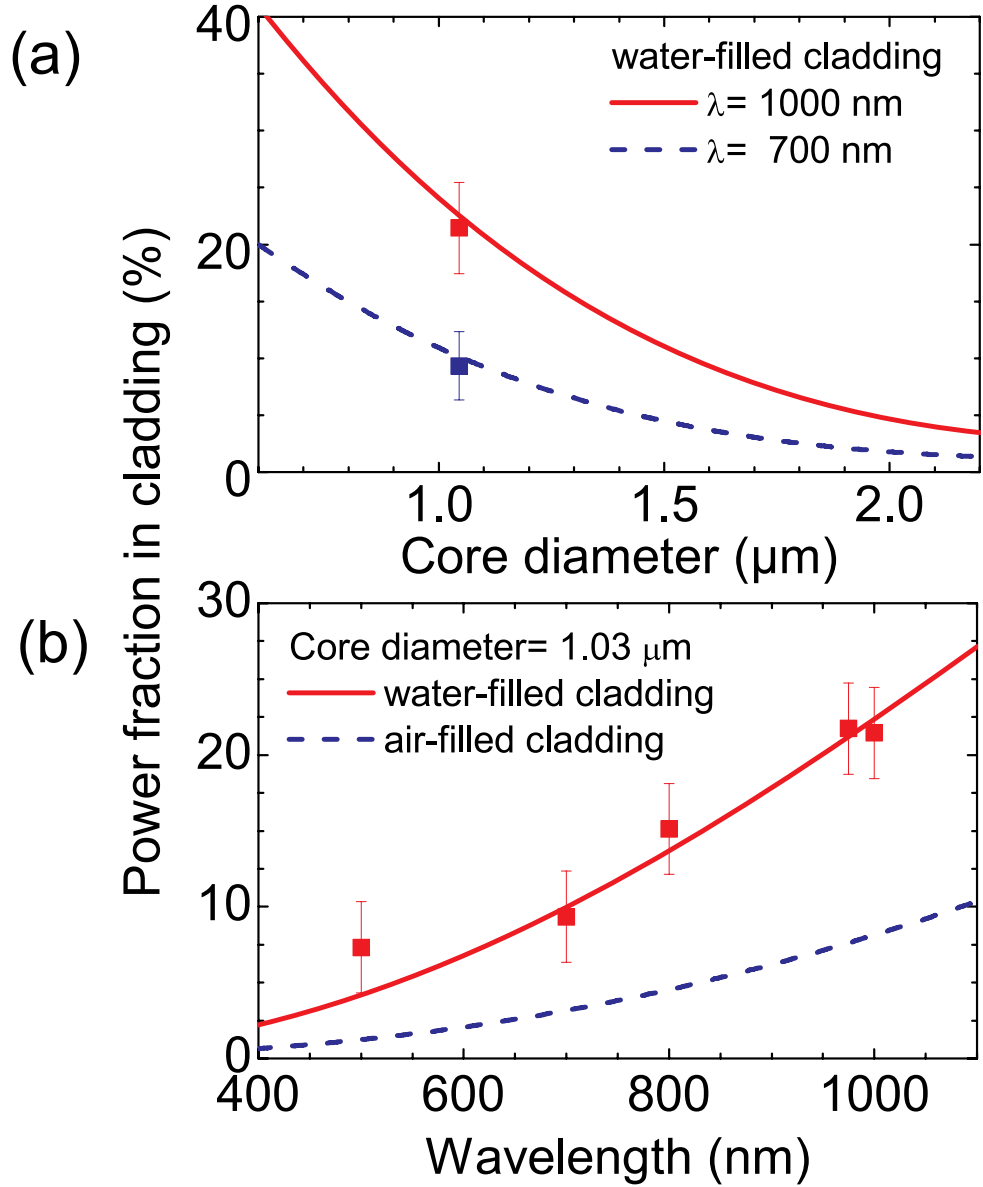}\\
\caption{\label{fig-powfrac}(Color online) (a) Calculated cladding power fraction $\Phi$ in a H$_2$O-filled fiber versus $d_{\mathrm{eff}}$ (curves) at $\lambda$=700~nm and $\lambda$=1000~nm; $\Phi$ increases with decreasing $d_{\mathrm{eff}}$. The symbols are power fractions for Fiber 2 that were obtained from the measured mode profiles in Figs.~\ref{fig-mode-profile}(a) and (c). (b) Calculated $\Phi$ versus wavelength for Fiber 2 with both H$_2$O- and air-filled cladding (curves). Both curves show that $\Phi$ increases with wavelength. The data points are experimental values obtained from measured mode profiles and are in quantitative agreement with theory (within 3\%).}
\end{center}
\end{figure}

The dependence of the calculated power fraction in the cladding holes on $d_{\mathrm{eff}}$, cladding medium and wavelength of the propagating light in the core is shown in Fig.~\ref{fig-powfrac}. It increases with wavelength, as light with longer wavelengths is less tightly confined to the solid core. By inserting an aqueous sample into the holes, the field extends further into the cladding due to the decreased index contrast. The experimental power fractions in the cladding were extracted from the measured mode profiles shown in Fig.~\ref{fig-mode-profile}. The beam profiles were multiplied with convolution masks generated from the SEM images. The optimum position and orientation of the masks were determined by an automated cross-correlation routine, in which the amount of light in the glass core was optimized. The squares in Fig.~\ref{fig-powfrac} show the resulting measured power fractions for a range of core sizes. We find quantitative agreement between measured and calculated power fractions to within 2\% over the entire wavelength range.

\section{DISPERSION}
\label{dispersion}

\begin{figure}[!ht]
\begin{center}
\includegraphics[width=0.8\linewidth]{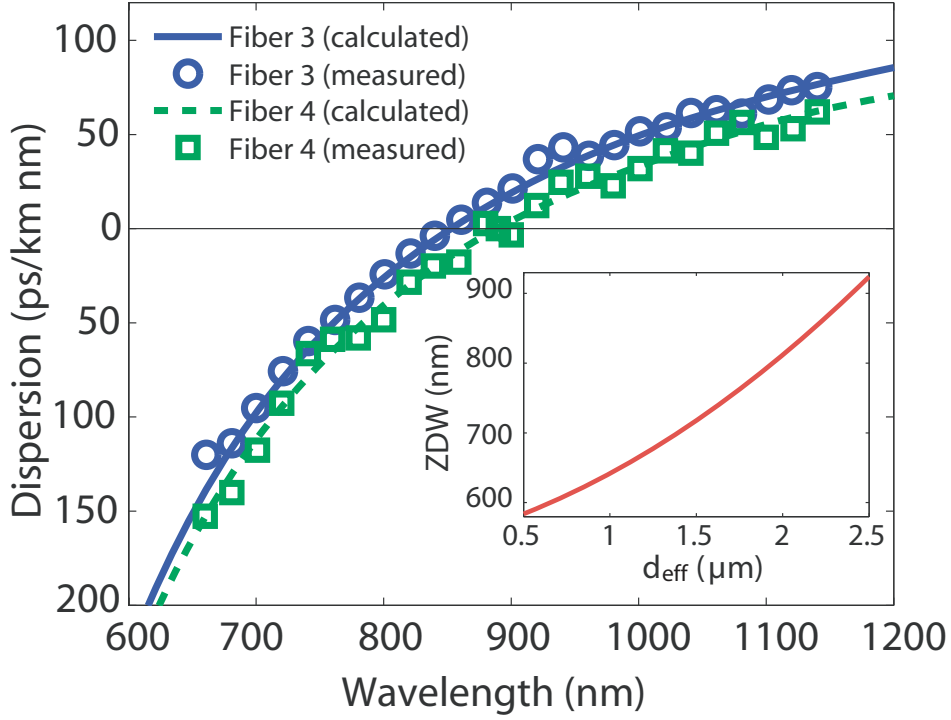}\\
\caption{\label{fig-dispersion}(Color online) Measured and calculated dispersion of Fiber 3 and Fiber 4 in the range between $\lambda$=600 and 1200~nm. The measured data for Fiber 3 (circles) and Fiber 4 (squares) show a zero dispersion wavelength (ZDW) of  $\lambda$=846~nm  and $\lambda$=887~nm, respectively. The solid curves represent dispersion curves obtained from FEM calculations without free parameters that agree within 2~ps$\cdot$nm$^{-1}$$\cdot$km$^{-1}$ with a polynomial fit (not shown) of the measured datapoints over the entire wavelength range. The inset shows the dependence of the calculated ZDW on $d_{\mathrm{eff}}$.}
\end{center}
\end{figure}

Another important characteristic of the fiber is the dispersion (Fig.~\ref{fig-dispersion}). The dispersion for Fibers 3 (circles) and 4 (squares) was measured with white light interferometry\cite{rashleigh1978} using SC generated from an ESM-PCF. We compare our measurements to FEM calculations taking into account the dispersion of silica. The calculated dispersion of Fibers 3 and 4 between $\lambda$=600 and 1200~nm is shown as the solid and dashed curves in Fig.~\ref{fig-dispersion}, respectively. The calculated dispersion of both fibers agrees within 2~ps$\cdot$nm$^{-1}$$\cdot$km$^{-1}$ with a polynomial fit through the measured datapoints over the entire wavelength range. The excellent agreement between our measurements and the theory is remarkable since no parameters were freely adjustable, exemplifying the accuracy of our FEM calculations.

The region near the zero dispersion wavelength (ZDW) is particularly interesting for nonlinear optical experiments.  Fibers 3 and 4 show ZDWs at  $\lambda$=846~nm and $\lambda$=887~nm, respectively. We can tailor the ZDW for nonlinear experiments in either the solid core of the fiber, or in the cladding holes. The ZDW of the fibers is controlled only by the $d_{\mathrm{eff}}$. The inset of Fig.~\ref{fig-dispersion} shows how the calculated ZDW depends on $d_{\mathrm{eff}}$. The first ZDW shifts towards the blue as the core size decreases, in agreement with the silica strand model.\cite{birks1999}

Recently, SC generation was demonstrated by launching regeneratively amplified Ti:sapphire pulses into a 10~cm long air-suspended solid-core fiber.\cite{cordeiro2007} While the generated supercontinuum was 730~nm broad, the required pulse intensity exceeded 1~TWcm$^{-2}$ (assuming a conservatively estimated coupling efficiency of 1\%). The high peak intensity and short fiber length suggest that the ZDW did not play a dominant role in the experiment. Clearly, by carefully tuning the ZDW to lie close to the pump wavelength the required pulse intensity for SC generation can be dramatically reduced, allowing the use of standard Ti:sapphire or microchip lasers as pump sources. We propose that such efficient SC generation could be used in single-fiber sensors in which both SC source and sample chamber are combined.

\section{IDEAL SENSING CONDITIONS}
\label{ideal-sensing}

An important goal in absorption spectroscopy is the identification of sample materials by the spectral shape of their absorption. Furthermore, the concentration of  certain chemical compounds in the sample can be deduced from the magnitude of the corresponding absorption peak. The absorption of light by a sample is commonly described by the Lambert-Beer law relating the absorption of light to the properties of the material through which it is traveling. Here, we use a modified version of this law taking into account the fraction of the light that travels through the sample $\Phi$($\lambda$). The resulting absorbance is:

\begin{equation}\label{eqn-abs}
    A(\lambda) [\text{dB}]=10\log_{10} \left( \frac{P_0}{P_1} \right)=\epsilon(\lambda) C  \ell \Phi(\lambda),
\end{equation}

\noindent where $P_0$ is the power of the incident light, $P_1$ is the transmitted light, $\epsilon$($\lambda$) is the molar absorptivity of the sample in Lmol$^{-1}$cm$^{-1}$, $C$ is the molar concentration of the sample in mol/L, and $\ell$ is the length of the sample length in cm.\cite{alpha} The spectral shape of the absorption thus depends both both on the sample property $\epsilon$($\lambda$) and on the fiber property $\Phi$($\lambda$) (see Fig.~\ref{fig-powfrac}). The magnitude of the absorption peak depends on the concentration, and fiber length.

We have constructed a general parameter diagram to obtain further insight into the optimum design parameters for hollow fiber sensors. In the diagram shown in Fig.~\ref{fig-param-space}, the absorbance is kept constant at\emph{ A}=5~dB, which is sufficiently large to be detected by any spectrometer. The dashed lines in the $\Phi$-$\ell$ plane indicate contours of constant $\epsilon$$C$ that were obtained from Eq.~\ref{eqn-abs}. For a given $\epsilon$$C$, any combination of $\Phi$ and $\ell$ that lies on the corresponding line will result in a 5~dB absorbance signal. The dashed lines thus define regions in which optimum sensing conditions can be achieved. The circle corresponds to the experiment shown in Fig.~\ref{fig-nicl-abs}, and lies on the contour for $\epsilon$$C$=0.0412~cm$^{-1}$ (solid line). The ideal sensing region is bounded by two factors. Firstly, for lengths longer than 20~m, fiber losses become considerable, reducing the amount of light that reaches the spectrometer. Secondly, we assume a practical upper limit of $\Phi$=50\%, based on our current experiments. Furthermore, for high values of $\Phi$, the slope of $\Phi$($\lambda$) becomes large (see Fig.~\ref{fig-powfrac}(b)), resulting in a sensitivity gradient in the measured absorption spectra. While the gradient in $\Phi$ can be compensated for, it is preferable to operate at lower $\Phi$ values, where $\Phi$ does not vary much with wavelength. Within these limits, the smallest detectible absorbance is as low as $\epsilon$$C$=0.0005~cm$^{-1}$. Such a high sensitivity can be achieved by choosing a fiber length of 20~m and $\Phi$=50\%. For comparison, we plot the data for a standard 1~cm long cuvette on the same diagram (red square in the bottom right corner of Fig.~\ref{fig-param-space}); the $\epsilon$$C$ value required for 5~dB is 0.5~cm$^{-1}$. From our analysis we conclude that air-suspended solid-core fibers can provide an increase in sensitivity of three orders of magnitude compared to standard cuvette measurements. While we have focused on liquid-filled fiber sensors, our diagram also applies to gas sensors.

\section{CHEMICAL SENSING}

Here we demonstrate broadband chemical sensing of an aqueous NiCl$_{2}$ solution (in which nickel is present largely as [Ni(H$_2$O)$_6$]$^{2+}$). NiCl$_{2}$ is a compound that is commonly used for electroplating and also in batteries. NiCl$_2$ is hazardous for the environment and particularly toxic to aquatic organisms. The LC$_{50}$/96-hours\cite{LC50} for water organisms is about 100~mg/L, corresponding to $C$=4.2$\times$10$^{-4}$~mol/L. Unfortunately, efficient monitoring of NiCl$_{2}$ concentrations is hampered by the low molar absorptivity of [Ni(H$_2$O)$_6$]$^{2+}$. Thus NiCl$_2$ is a compound highly suitable for testing the performance of our fiber sensor.

\begin{figure}[!ht]
\begin{center}
\includegraphics[width=0.9\linewidth]{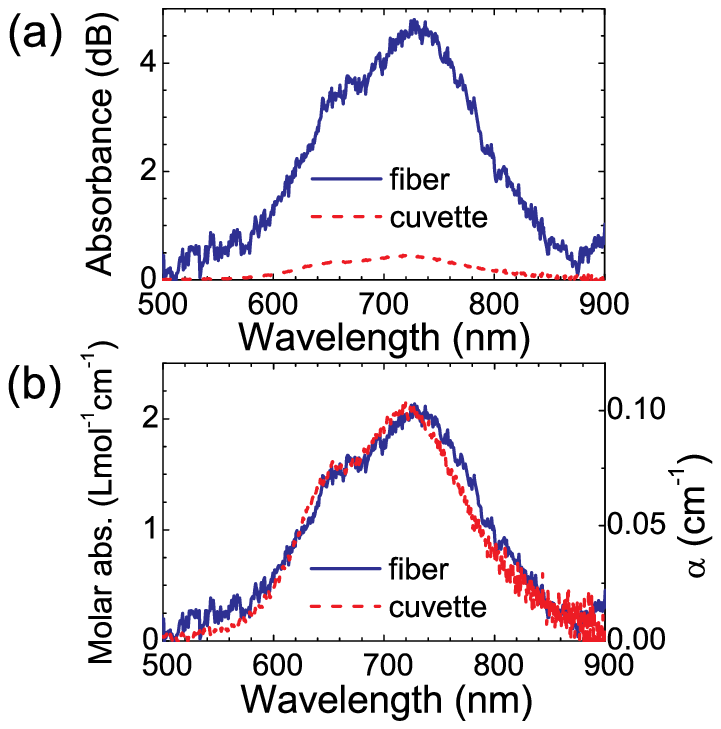}\\
\caption{\label{fig-nicl-abs}(Color online) (a) Absorbance spectrum of an aqueous 2.1$\times$10$^{-2}$~mol/L NiCl$_2$ solution, normalized to H$_2$O reference, measured in a $\ell$=1.1~m long piece of Fiber 2 (solid) and in a 1~cm standard cuvette (dashed). In both spectra we observe two sub-peaks at $\lambda$=720~nm and $\lambda$=660~nm. In the fiber, however, the absorbance strongly increases from 0.4 to 4.7~dB. (b) Molar absorptivity spectrum obtained by dividing the absorbance data in (a) by $\ell$$\Phi$$C$. The right-hand axis shows the corresponding absorption coefficient $\alpha$. The excellent agreement is striking since no parameters were freely adjusted.}
\end{center}
\end{figure}

To compare our measurement to standard spectroscopic techniques, we chose a NiCl$_2$ concentration $C$=2.1$\times$10$^{-2}$~mol/L that is just detectible in a $\ell$=1~cm cuvette measurement. To obtain the  $\epsilon$($\lambda$) reference spectrum, we measured the transmission of a collimated halogen light source through a 1~cm long cuvette. The transmitted spectrum was measured with an Ocean Optics HR-4000 spectrometer. The dashed curve in Fig.~\ref{fig-nicl-abs}(a) shows the absorbance in the 500 to 900~nm range, obtained by normalizing to the transmission through an H$_2$O-filled cuvette. The absorbance reaches a maximum value of 0.4~dB at a wavelength of 720~nm. The dashed curve in Fig.~\ref{fig-nicl-abs}(b) shows the resulting molar absorptivity spectrum.

\begin{figure}[!ht]
\begin{center}
\includegraphics[width=0.9\linewidth]{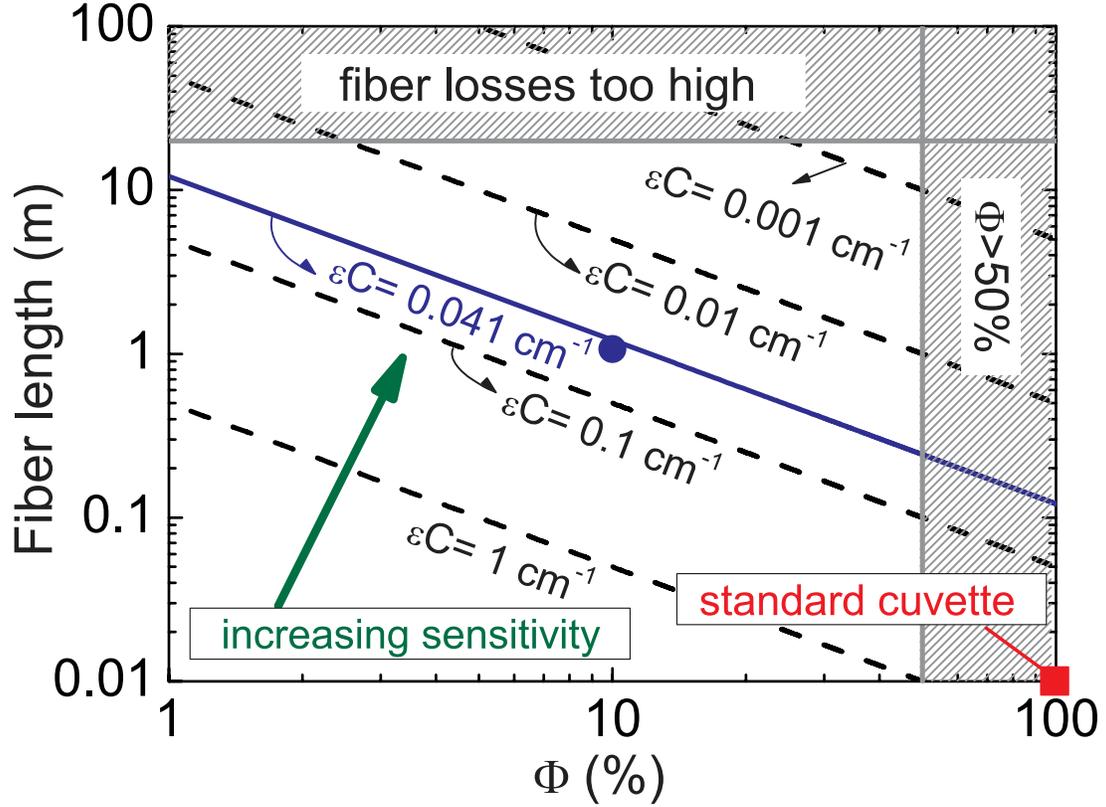}\\
\caption{\label{fig-param-space}
(Color online) Diagram mapping out the ideal sensing conditions. Contours of constant absorbance are plotted in the $\Phi$-$\ell$ plane, defining regions in which optimum sensing conditions can be achieved. The absorbance is kept constant at \emph{A}=5~dB, sufficiently large to be detected by any spectrometer. The dashed lines are contours of fixed $\epsilon$$C$. For a given $\epsilon$$C$, any combination of $\Phi$ and $\ell$ that lies on the corresponding line will result in a 5~dB absorbance signal. The circle corresponds to the experiment shown in Fig.~\ref{fig-nicl-abs}, and lies on the contour for $\epsilon$$C$=0.0412~cm$^{-1}$ (solid line). The green arrow indicates the direction in which the sensing sensitivity increases. The ideal sensing region is bounded by two factors. Firstly, for lengths longer than 20~m fiber losses are considerable, reducing the amount of light reaching the spectrometer. Secondly, we assume a practical upper limit of $\Phi$=50\%, based on our own experience. Given these boundaries, the smallest detectible absorbance is as low as $\epsilon$$C$=0.0005~cm$^{-1}$. The red square in the bottom right corner represents a standard 1~cm long cuvette, and corresponds to $\epsilon$$C$=0.5~cm$^{-1}$.}\end{center}
\end{figure}

The absorption spectrum of [Ni(H$_2$O)$_6$]$^{2+}$ is known to exhibit three broad absorption bands between 350~nm and 1400~nm, arising from spin-allowed d-d electronic transitions. The central absorbance band, which we observed to be split to give two maxima at 720~nm and 660~nm (in accordance with the literature values of 720~nm and 656~nm \cite{Egghart69,Bussiere98}) exhibited molar absorptivities of 2.1~Lmol$^{-1}$cm$^{-1}$ and 1.5~Lmol$^{-1}$cm$^{-1}$ respectively. This central absorbance corresponds to the $^\textrm{3}$A$_{\textrm{2g}} \rightarrow _\textrm{3}$T$_{\textrm{1g}}$(F) electronic transition. In the absence of coupling this would be expected to give a single maximum. However, due to the presence of strongly coupled electronic states (in this case $^\textrm{3}$T$_{\textrm{1g}}$ and $^\textrm{1}$E$_{\textrm{g}}$), a superposition of several transitions is detected giving rise to the two maxima observed.\cite{Triest00} The resulting value of $\epsilon$\emph{C } for a 2.1$\times$10$^{-2}$~mol/L NiCl$_2$ solution at $\lambda$=720~nm is 0.041~cm$^{-1}$, and corresponds to the solid curve in the ideal-sensing diagram in Fig.~\ref{fig-param-space}.

For the fiber-based measurement, we chose Fiber 2 with $d_{\mathrm{eff}}$=1.05~$\mu$m, which has a cladding power fraction of $\Phi$=10.4\% at $\lambda$=700~nm --- the center of the absorption band for NiCl$_2$. Fig.~\ref{fig-param-space} shows that the optimum fiber length required for an \emph{A}=5~dB absorbance signal is $\ell$=1.1~m (displayed as a circle in Fig.~\ref{fig-param-space}). The fiber was connected to water cells and filled first with H$_2$O to obtain a reference spectrum. The sample volume in the fiber (1~$\mu$L) is reduced by three orders of magnitude compared the cuvette measurement (1~mL). The transmission of a broadband SC source was measured using an OSA (as described in Section~\ref{tx-loss}). Subsequently, the H$_2$O in the fiber was replaced by an aqueous 2.1$\times$10$^{-2}$~mol/L NiCl$_{2}$ solution. The solid curve in Fig.~\ref{fig-nicl-abs} shows the resulting absorbance spectrum, obtained by normalizing the NiCl$_2$ data to the transmission of the same fiber filled with H$_2$O. We observe the same broad absorption band between $\lambda$=600 and 800~nm as in the cuvette. Importantly, our fiber spectrum also resolves the two sub-peaks at $\lambda$=720~nm and $\lambda$=660~nm, illustrating that the fiber does not introduce spectral artefacts. The maximum measured absorbance \emph{A}=4.7~dB is in good agreement with our prediction. A direct quantitative comparison between our fiber data and the reference sample is made in Fig.~\ref{fig-nicl-abs}(b). Here, we extract the molar absorptivity spectrum $\epsilon$($\lambda$) of NiCl$_2$ from our data by dividing the absorbance data by $\ell$$\Phi$$C$, where we have used the cladding power fraction $\Phi$=10.4\% at $\lambda$=700~nm. We find striking agreement between our in-fiber measurement (solid curve) and the reference spectra measured in a standard cuvette (dashed curve). Our results demonstrate that air-suspended solid-core fibers can be used in quantitative broadband chemical sensing measurements.

\section{CONCLUSIONS}
\label{conclusion}

Air-suspended solid-core PCFs can be used to make quantitative broadband measurements via evanescent-field sensing. Measured mode profiles are in good agreement with FEM calculations without free parameters, and the cladding power fraction can be tuned via the core diameter of the fiber. Dispersion measurements, in excellent agreement with theory, demonstrate tuning of the ZDW via the core diameter. Optimum design parameters for absorption sensors can be summarized in a general diagram. We estimate that a sensitivity increase of three orders of magnitude (compared to standard cuvettes) is feasible for both liquids and gases. In measurements of the broad absorption peak in an aqueous NiCl$_2$ solution, striking agreement is obtained with the reference spectrum measured in a standard cuvette, even though the sample volume is decreased by three orders of magnitude. Our results demonstrate that air-suspended solid-core PCFs can be used in quantitative broadband chemical sensing. This opens up new opportunities both for environmental sensing and in quantitative chemical analysis. Experiments in which the cladding holes are used to study reaction dynamics of confined chemical processes are currently underway.

\section{ACKNOWLEDGMENTS}
We thank Andre Brenn and Gustavo Wiederhecker for discussions. We acknowledge the K\"orber Foundation for funding this research project.

\newpage

\begin{thebibliography}{}
\bibitem{alpha}
We use a base-10 logarithmic definition for the absorbance \emph{A}. The base-$\ln$ absorption coefficient $\alpha$ is easily obtained from $\epsilon$$C$ by multiplying by a prefactor: $\alpha$=$\epsilon$$C$/$\log_{10}(e)$.
\end{thebibliography}
\end{document}